\documentclass[aps,prb,twocolumn,showpacs,superscriptaddress,floatfix,amsmath,amssymb,amsfonts]{revtex4-1}
\usepackage{graphicx}
\usepackage{dcolumn}
\usepackage{bm}
\usepackage{amsmath}
\begin{document}
\title{Structural, electronic and magnetic properties of Fe doped CoCr$_{2}$O$_{4}$: insights from ab initio calculations}
\author{Debashish Das} %
\affiliation{Department of Physics, Indian Institute of Technology
Guwahati, Guwahati, Assam 781039, India} %
\author{Shreemoyee Ganguly}
\affiliation{S.N.Bose National Center for Basic Sciences,
JD Block Sector III, Salt Lake City, Kolkata 700098, India}

\author{Biplab Sanyal}
\affiliation{Department of Physics and Astronomy,
Uppsala University, Box 516, 75120 Uppsala, Sweden}
\email{For correspondence: Biplab.Sanyal@physics.uu.se}

\author{ Subhradip Ghosh}
\affiliation{Department of Physics, Indian Institute of Technology
Guwahati, Guwahati, Assam 781039, India} %

\date{\today}

\begin{abstract}
 CoCr$_2$O$_4$ has attracted significant attention recently due to several interesting properties such as magnetostriction, 
magnetoelectricity etc.. More recent experiments on Fe substituted CoCr$_2$O$_4$ observed a variety of novel
phenomena such as the magnetic compensation accompanied by the occurrence of exchange bias, which reverses its sign. 
Understanding of such phenomena may lead to control the properties of these material in an efficient way to enhance its potential for 
multifunctional applications. In this paper, we study the microscopic understanding of Fe doping in
modifying the structural and magnetic properties of CoCr$_{2}$O$_{4}$ with varying composition and substitution of Fe at different sublattices
by first-principles density functional calculations. We have analysed in detail the effect of Fe substitution on crystal field and exchange splittings,
magnetic moments and interatomic exchange parameters. It is also observed that with increasing concentration of Fe impurity, the
system has a tendency towards forming an "Inverse Spinel" structure as observed in experiments. Such tendencies are crucial to understand this system as 
it would lead to modifications in the magnetic exchange interactions associated with sites with different symmetry finally 
affecting the magnetic structure and the multiferrocity in turn.
\end{abstract}

\maketitle
\section{Introduction}
CoCr$_{2}$O$_{4}$, a member of the spinel oxide family, has received considerable attention in past
few years after the discovery of magnetisation reversal of ferroelectric polarisation in this material
\cite{prl06}. Subsequent studies, in order to explore the multiferroic nature of this system, resulted in
illuminating temperature and magnetic field dependent magnetocaloric properties \cite{prl09,apl11}. This
material has a rich magnetic phase diagram. At a temperature around 93 K, it exhibits a collinear
ferromagnetic ordering followed by a transformation to a complex incommensurate conical spin spiral around 26 K
\cite{jpf64,prb04}. This conical spin spiral further transforms to a commensurate one \cite{jpcm09}, which
further undergoes unconventional magneto-structural transitions at high magnetic fields \cite{prb12,prl13}.
The origin of ferroelectric polarisation in this material is believed to be due to the conical spin spiral state
and is explained by the inverse Dzyaloshinskii-Moriya effect \cite{prl06}. 

CoCr$_{2}$O$_{4}$ crystallises in normal spinel structure in which the Co$^{2+}$ ions occupy the A sites having 
tetrahedral symmetry and the Cr$^{3+}$ ions occupy the B sites with octahedral symmetry. The magnetoelectric 
coupling in this material is 
attributed to the conical order of the Cr$^{3+}$ sublattice \cite{prl06}. This very well fits into the so called
LKDM theory \cite{lkdm} which, by using a simple Heisenberg model, demonstrates that the magnetic exchange
interactions among A-B and B-B ions determine the magnetic order in cubic spinels. However, recent measurements
\cite{jpcm09,jpsj07} and first-principles electronic structure calculations \cite{ederer} establish that the
magnetic structure at the A site and the magnetic exchange interactions among A ions cannot be neglected at
low temperatures. Therefore, the variations in the occupancies at the A and B sub-lattices can induce significant
changes in the magnetic interactions, thus influencing the magnetic order and the associated electric polarisation.


It is, thus, interesting to investigate the effects of substitutions by a third magnetic cation in CoCr$_{2}$O$_{4}$,
as that might affect the magnetic exchange interactions, thereby affecting the stabilities of the spin spiral states leading to interesting functional properties. To this end, recent attempts on Fe substitution at the B sublattice have been remarkable.
Magnetic measurements on Co(Cr$_{0.95}$Fe$_{0.05})_{2}$O$_{4}$ \cite{padamapl} and on Co(Cr$_{0.925}$Fe$_{0.075})_{2}$O$_{4}$
\cite{padamaip} show the phenomenon of magnetic compensation and sign reversal of Exchange
Bias \cite{eb} around a critical temperature.
These unusual phenomena vanish when Fe content reaches about $12.5 \%$ in the B sublattice. These effects were attributed
to the spin reorientations of the magnetic cations at various sublattices \cite{padamapl,padamthesis}. This point was
further elucidated by Zhang {\it et al} \cite{pss13} who upon investigations of the magnetic compensations in
CoCr$_{2-x}$Fe$_{x}$O$_{4}$ systems inferred that the magnetic compensation can be understood in terms of a role
conversion of magnetic contributors and a composition compensation between two competitive magnetic sublattices at
$x=0.1$, the origin of which is the changes in the sublattice occupancies of the Fe$^{3+}$ ions. They also found a
significant magnetostriction at $x=0.4$, which is consistent with the composition and temperature dependent compensation
behaviour.

These experimental results demonstrate that the understanding of the competitive roles of the magnetic components at 
various sublattices is the key to comprehend the fascinating physics exhibited by Fe substituted CoCr$_{2}$O$_{4}$. Some of us have recently studied
\cite{spinelpap1} the evolution of magnetic and electronic structures in CoCr$_{2}$O$_{4}$ while going from 'normal spinel' to 'inverse spinel' by Fe doping. 
However, the structural aspects and their connection to electronic and magnetic structures was not considered in that study.
In this communication,
we, report density functional theory (DFT) based calculations on the structural, electronic and magnetic properties of Fe doped CoCr$_{2}$O$_{4}$
with a specific focus on doping at different sublattices. Our results throw light on the fundamental aspects of structural deformations, the comparative roles of crystal 
field and magnetic exchange fields, the electronic correlations, doping at different sublattices and the hybridisations
due to Fe substitution in CoCr$_{2}$O$_{4}$. These results can help in microscopic understanding of the physical interactions in
this system. The paper is organised as follows: Section II details the methodology used, section III is dedicated
to the results and discussions followed by the conclusions. 
\section{methodology}
CoCr$_{2}$O$_{4}$ crystallises in the spinel structure with space group
$Fd \bar{3}m$.The structure can be described as an approximately close-packed arrangement of oxygen anions with metal cations 
distributed amongst the tetrahedral and octahedral sites in A and B sublattices respectively. The eightfold positions (000 : $\frac{1}{4} \frac{1}{4} \frac{1}{4})$
are the tetrahedral sites occupied by the Co$^{2+}$ cations, while the sixteen fold positions
$(\frac{5}{8} \frac{5}{8} \frac{5}{8}: \frac{5}{8} \frac{7}{8} \frac{7}{8}: \frac{7}{8} \frac{5}{8} \frac{7}{8}: \frac{7}{8} \frac{7}{8} \frac{5}{8})$
are the octahedral sites occupied by the Cr$^{3+}$ cations. The oxygen anion sublattice is arranged in a pseudo-cubic close packed spatial arrangement. 
A 14 atom unit cell (2 formula units) with $Fd \bar{3}m$ symmetry
is considered for pristine CoCr$_{2}$O$_{4}$ in the present calculations. 
Collinear Neel configuration is considered as the magnetic structure where all Cr spins are parallel to each other but are 
antiparallel to the Co spins.To model Fe substituted CoCr$_{2}$O$_{4}$ we have considered a $2\times2\times2$ supercell 
of the 14 atom primitive cell consisting of 112 atoms. 
We have investigated the systematic effects of Fe doping by
first replacing one Cr atom and then two Cr atoms. Substitution of one Cr atom leads to Co(Cr$_{1-x}$Fe$_{x}$)$_{2}$O$_{4}$
with $x \sim 0.03125$ amounting to 6.25 $\%$ Fe content at B sites, while substitution of two Cr atoms lead to the
case with $x \sim 0.0625$ amounting to 12.5 $\%$ Fe content at B sites. It is to be noted that these $x$ values are close to 
$x=0.05$ where the phenomena like magnetic compensation, negative magnetisation and reversible EB effect have been observed
\cite{padamapl}.

The total energies, structural parameters, electronic structure and magnetic properties are calculated by DFT+U \cite{dftu}
method. Inclusion of the effects of strong correlations is a necessity for oxides \cite{ox1,ox2,ox3,ox4} and one of the simplest methods is to use
DFT+U. Also, it was shown \cite{ederer} that non-incorporation of correlation effects lead to wrong ground state for
CoCr$_{2}$O$_{4}$. In this work, we have used the DFT+U approach
adopted by Dudarev {\it et al} \cite{dudarev}. In this approach, the effect of correlation is addressed through the effective
Hubbard parameter $U_{eff}=U-J$, where $U$ denotes the strength of the Coulomb interaction, the so called "Hubbard U" and $J$ is
the intra-atomic Hund's parameter. We have used 5 eV for $U_{Co}$, 3 eV for $U_{Cr}$ and 4 eV for $U_{Fe}$. The value of $J$
was taken to be 1 eV for all the elements. Such choices were based upon the available information \cite{u1,u2,u3,ederer}. We have used
the Projector Augmented Wave (PAW) \cite{paw} formalism as implemented in VASP code \cite{vasp}. The exchange-correlation part of the
Hamiltonian was treated with PBE-GGA functional \cite{pbe}.
A plane wave cut-off energy of 400 eV was considered for all the cases. The positions of all atoms
were relaxed keeping the volume fixed at the experimental one. The relaxations were carried out till the forces are less than 
$10^{-3}$ eV/\AA. The electronic self-consistency cycle is iterated till the total energy is converged better than $10^{-5}$ eV. 
A $2 \times 2 \times 2$ $\Gamma$ centred $k$-mesh was used for self-consistency and a $5 \times 5 \times 5$ mesh was used for
calculations of the electronic structure.  

The structural distortions in the compounds are assessed by computing angles $\theta_{T_{d}}$ and $\theta_{O_{h}}$, which denote the distortions from perfect
tetrahedral and octahedral symmetries respectively, and are defined as,
\begin{eqnarray}
\theta_{T_d}&=&\Sigma_{i=1}^{4} \frac{|109.5-\theta^{T_d}_i|}{n} \nonumber \\
\theta_{O_h}&=&\Sigma_{i=1}^{8} \frac{|90-\theta^{O_h}_i|}{n}
\label{thetatdoh}
\end{eqnarray}  
$\theta_{i}^{T_{d}},\theta_{i}^{O_{h}}$ are the ligand-metal-ligand bond angles 
corresponding to the $i$-th bond for the tetrahedral and octahedral symmetries
respectively.

The magnetic properties are investigated by computing the total and partial magnetic moments and the magnetic exchange parameters. To
determine the magnetic exchange parameters $J_{ij}$ among sub-lattices $i,j$,
 we have calculated total energies for different collinear
magnetic configurations and mapped them onto a classical 
Heisenberg Hamiltonian
\[E=-\sum_{i,j}J_{ij}\vec{S_i}.\vec{S_j}\]
where $S_i$ and $S_j$  are the unit spin vectors. It should be noted that $J_{ij}>$0 ($<$0) denotes ferromagnetic (antiferromagnetic) coupling. 

\section{Results and Discussions}
\subsection{Pristine CoCr$_{2}$O$_{4}$}
Before embarking on the effect of Fe doping, the pristine CoCr$_{2}$O$_{4}$ is investigated, in particular the structural
aspects. We have done calculations at the experimental lattice constant of 8.335 \AA \cite{expcocr2o4} and optimised the
internal structure parameter $u$, which corresponds to the oxygen sites. Our calculated value of $u$ is 0.261, which agrees very well
with the earlier results \cite{ederer,expcocr2o4}. As was mentioned earlier, the magnetic configuration considered is the collinear
Neel configuration. We have considered the Cr magnetic moments to be "spin up" and the Co moments to be "spin down". Our calculations
showed both Co and Cr in high spin states with magnetic moments -2.66 $\mu_{B}$ and 2.93 $\mu_{B}$ respectively, resulting in a 
magnetic moment of about 3 $\mu_{B}$ per formula unit for the system. 

\begin{figure}[ht]
\includegraphics[scale=0.7,  keepaspectratio]{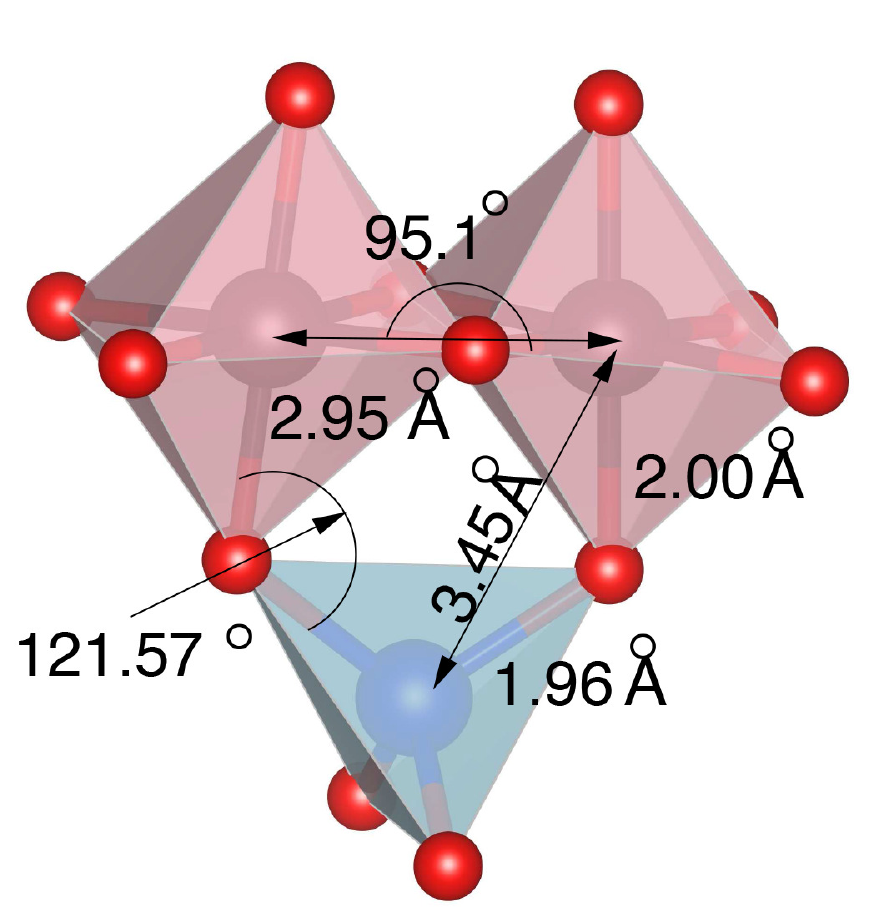}
\caption{\label{fig:pure} A part of near neighbourhood of the tetrahedrally coordinated $Co$ atom and octahedrally coordinated
$Cr$ atom in pristine $CoCr_2O_4$ is shown. Various bond angles and bond lengths 
are shown. The Co atoms are shaded blue, the Cr atoms pink and the O atoms red. 
}
\end{figure}

In Fig. \ref{fig:pure} we show a part of the neighbourhood of the octahedrally coordinated Cr and tetrahedrally coordinated 
Co atom. All the relevant angles and bond lengths are also demonstrated. In order to assess the structural distortions in this 
compound, we have estimated the values of angles $\theta_{T_{d}}$ and $\theta_{O_{h}}$ in accordance with 
Eqn.(\ref{thetatdoh}). While $\theta_{T_{d}}\sim 0.03^{0}$ [the angles vary between $102.56^{0}$ and $109.48^{0}$] , $\theta_{O_{h}} \sim 5.34^{0}$ 
[the angles vary between 84.67 and 95.34]. The results suggest that the distortions associated with the tetrahedral sites are negligible.

\subsection{Co(Cr$_{0.96875}$Fe$_{0.03125}$)$_{2}$O$_{4}$}
We have mentioned earlier that the above composition is modelled by replacing one Cr atom by an Fe atom in the 112 atom
cell. This Fe atom can occupy a B site or can occupy an A site. In the later case, it will displace one Co atom which will occupy the
B site, vacant due to replacement of Cr. This substitution by Fe at A site mimics the tendency towards formation of "Inverse"
structure, where in the case of an AB$_{2}$O$_{4}$ compound, one B atom will occupy the tetrahedral site, pushing the lone
A atom to the octahedral site. The case of Fe substitution at A site is particularly interesting as it now provides us with,
apart from the substituting impurity at A site, the presence of Co atoms at both tetrahedral and octahedral sites. The
resulting effects on the structural distortions, the spin structure of the system and the electronic structure would throw
some light on the microscopic physics in the system which would become more relevant at higher Fe concentrations where the
inversion indeed starts to occur \cite{pss13}. In the following sub-sections we report results on both substitution at A site and
at B site.

\subsubsection{Fe substitution at A site}
We first report the minimum energy spin configuration for a single Fe atom substituting a Co at site A. In order to find this, we
have considered three different bond distances between 
Co at tetrahedral sites(Co$_{T}$) and Co at octahedral sites(Co$_{O}$). 
For each of the three
distances, we have calculated total energies of different spin orientations of atoms at A and at B sites to obtain the minimum
energy spin configuration. The results are summarised in Table \ref{tabone:table1}.
\begin{table}[ht]
    \caption{\label{tabone:table1}Total energy (meV) per Fe atom and the spin structure for different bond distances
between Co at tetrahedral site (Co$_{T}$) and Co at octahedral site (Co$_{O}$)
for Fe substitution at A site in CoCr$_{2}$O$_{4}$.}  
\begin{center}
        \begin{tabular}{|c|c|c|c|c|c|} \hline
    & \multicolumn{4}{|c|}{Orientation of spins} & Total energy \\ \hline
    Co$_{T}$-Co$_{O}$ bond & & & & & \\ 
    distance ($\AA$) & Co$_{T}$ & Co$_{O}$ & Cr & Fe & \\ \hline
    3.45 & $\downarrow$ & $\downarrow$ & $\uparrow$ & $\downarrow$ & 0 \\ \hline
    5.4 & $\downarrow$ & $\uparrow$ & $\uparrow$ & $\downarrow$ & 46 \\ \hline
    6.87 & $\downarrow$ & $\uparrow$ & $\uparrow$ & $\downarrow$ & 87 \\ \hline 
            \end{tabular}
       \end{center}
\end{table}
Our results show that the Co atoms at tetrahedral and at octahedral sites prefer to be nearest neighbours, the spins of
substituted Fe, and the Co at octahedral as well as as tetrahedral sites prefer to align anti-parallel to the spin of
the Cr atom. The results of magnetisation show that the substituted Fe atom is in a high spin state; so are the other atoms.
The magnetic moments of Co at tetrahedral site, the Co at octahedral site, the Fe and the Cr are -2.66 $\mu_{B}$, -2.71 $\mu_{B}$,
-4.02 $\mu_{B}$ and 2.93 $\mu_{B}$ respectively, yielding a total magnetic moment of about 2.5 $\mu_{B}$ per formula unit.
Thus, the moment of Co atoms remain almost constant with respect to those in pristine CoCr$_{2}$O$_{4}$ and the reduction
in the overall moment is due to the large anti-parallel moment of Fe.

After fixing the minimum energy spin configuration, we analyse the structural parameters and the electronic structure.
Since the lattice constant and oxygen $u$ parameter do not change appreciably for such a low concentration of Fe substitution \cite{pss13},
we have kept them the same as those of CoCr$_{2}$O$_{4}$ and relaxed the atom positions only.  It is observed that 
the oxygen atoms in the Co octahedra move outwards while those in the
Fe tetrahedra move inwards resulting in slight contractions in the substituted tetrahedra and slight expansions in the substituted
octahedra. This can be understood from the fact that the size of Co is larger than that of Cr and Fe. So, when the Co atom is substituted
at an octahedral site, it expands the octahedra locally while the Fe that substitutes a Co at tetrahedral site contracts the tetrahedra.

\begin{figure}[ht]
\includegraphics[scale=0.7,  keepaspectratio]{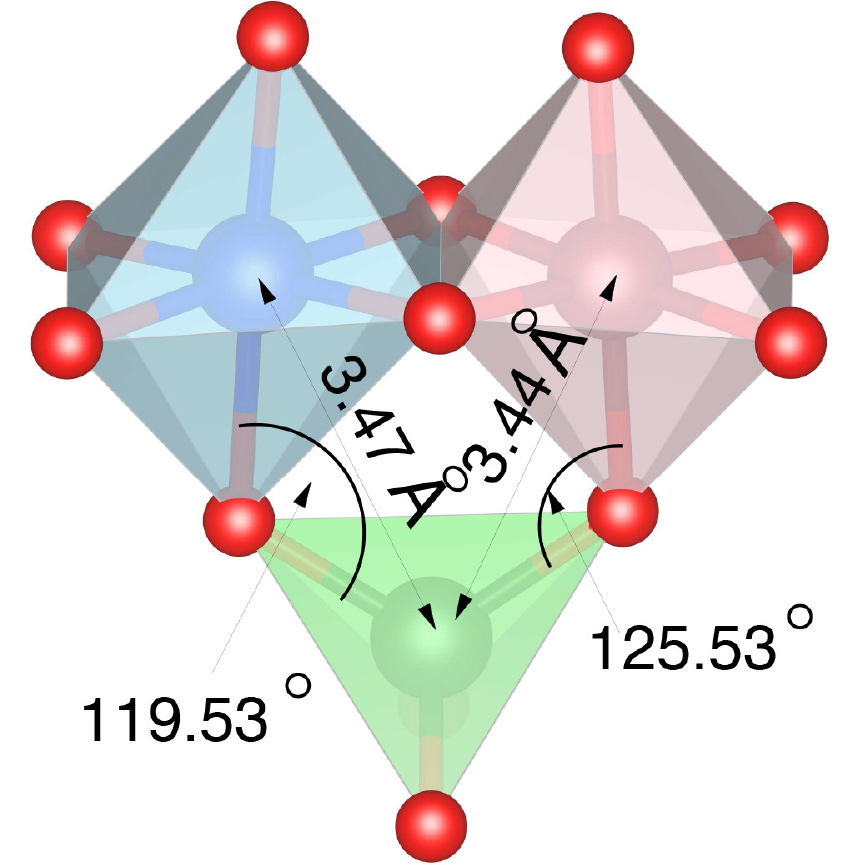}
\caption{\label{fig:a-site} A part of near neighbourhood of the tetrahedrally coordinated $Fe$ atom and octahedrally coordinated
$Co$ atom are shown along with various bond angles and bond lengths for single
Fe substituted at A site in CoCr$_{2}$O$_{4}$. 
The Co atoms are shaded blue, the Fe atoms
green, the Cr atoms pink and the O atoms red. 
}
\end{figure}

In Fig. \ref{fig:a-site} , we show the neighbourhood of the substituted transition metals along with the relevant structural 
parameters. Upon Fe
substitution at A site, the Fe-O-Cr and the Fe-O-Co$_{O}$ angles increase and decrease respectively in comparison to the Co-O-Cr
angle in CoCr$_{2}$O$_{4}$ quantifying the distortions around the substituted sites. The values of $\theta_{T_{d}}$ and $\theta_{O_{h}}$ [calculated using Eqn. (\ref{thetatdoh})] are $\sim 0.71^{0}$ [the angles vary between $108.36^{0}-112.32^{0}$] and $5.95^{0}$ [the angles vary between $83.80^{0}-96.95^{0}$] respectively. 
This implies that the distortion in the tetrahedral environment is more due to Fe substitution at tetrahedral sites. These results indicate that
the degeneracies in the $d$ orbitals of the transition metal atoms are going to be affected and one can expect a significant impact on
the electronic structure.


In what follows, we provide an understanding of the effects of structural distortion and electron-electron
correlation on the electronic structure and individual spin states of the constituents by a study of the
crystal field splitting (CF) and exchange splitting(Ex) through an analysis of the corresponding densities
of states. Before analysing the Fe substituted system, we first look at the pristine compound. The exchange
splitting of t$_{2g}$ and e$_{g}$ levels and the crystal field splitting of the spin channels for Co and Cr
occupying A and B sites respectively are listed in the first two rows of Table \ref{tab2}. These results are
obtained without the effects of correlations, that is, for $U=0$. The results suggest that in the down spin
channel ($\downarrow$) of A site, the five-fold 3d crystal field degeneracy is barely broken. Thus, this spin
band is almost completely occupied resulting in a high spin state for Co cation. The other 3d electrons
mostly occupy the exchange split ($\Delta^{e_{g}}_{EX}=1.5$ eV) e$_{g}$ up spin channel($\uparrow$). The
t$^{\uparrow}_{2g}$ states being energetically much higher ($\Delta^{t_{2g}}_{EX}=2.7$eV) remains unoccupied.
For the B site, the triply degenerate t$^{\uparrow}_{2g}$ states are energetically much lower than the 
corresponding e$_{g}^{\uparrow}$ ($\Delta^{\uparrow}_{CF}=3.2$eV) and t$^{\downarrow}_{2g}$ ($\Delta^{t_{2g}}_{EX}=3.2$ eV).
Thus these ($t_{2g}^{\uparrow}$)states are almost completely occupied resulting in a high spin state of Cr. The energy level diagram
of Figure 3(i) sums it up.
\begin{figure}[ht]
\includegraphics[width=8cm,  keepaspectratio]{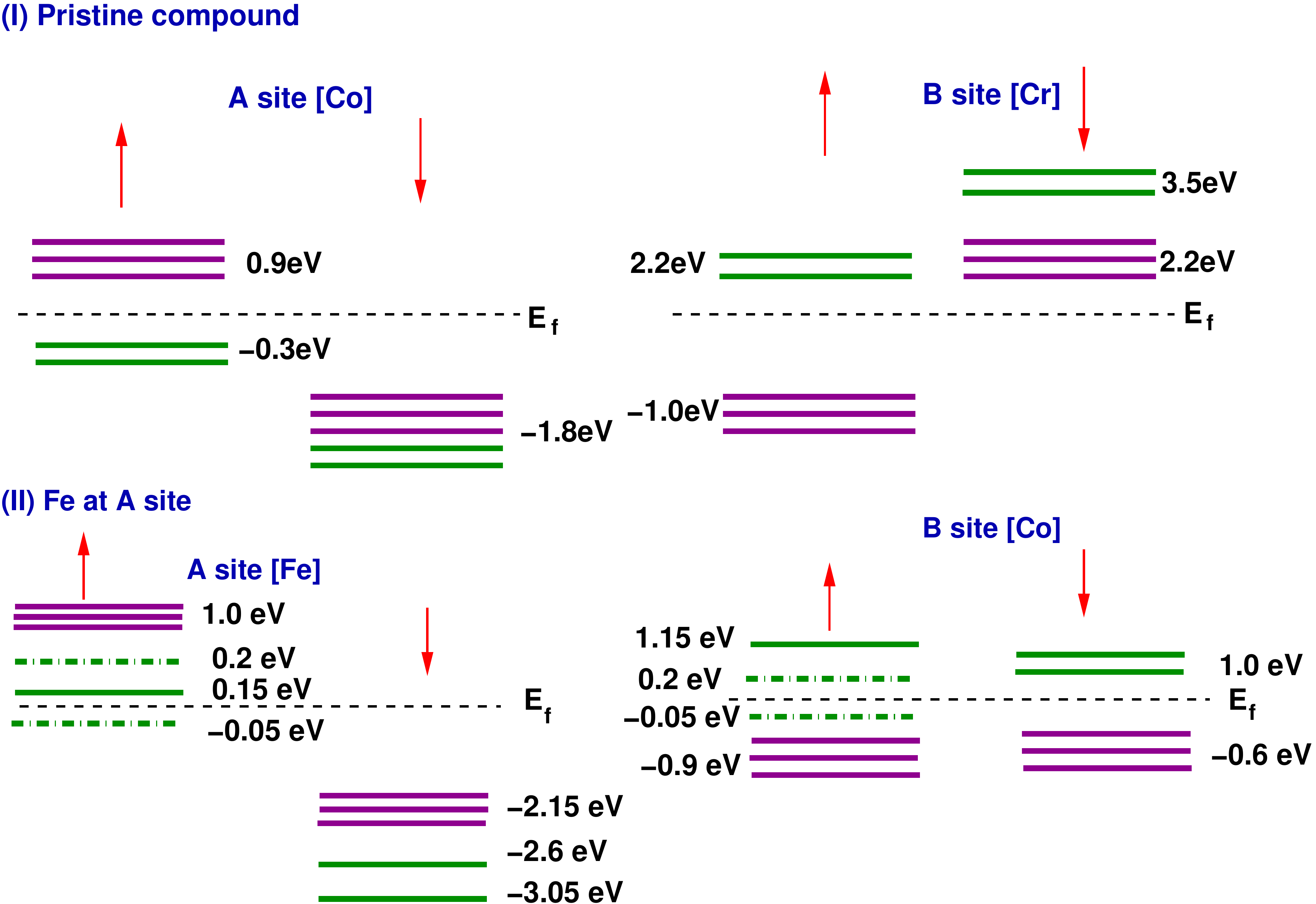}
\caption{\label{fig:energy-level-A}(i) The crystal field and exchange splitting of the t$_{2g}$($d_{xy}$,$d_{yz}$ 
and $d_{xz}$) and e$_{g}$ ($d_{z^2}$ and $d_{x^2-y^2}$) levels of pristine CoCr$_2$O$_4$ are analysed through Energy 
Level Diagram. While the t$_{2g}^{\downarrow}$, e$_{g}^{\downarrow}$ and e$_{g}^{\uparrow}$ are almost completely occupied for the tetrahedrally coordinated $\mathcal{A}$ site, only the
 t$_{2g}^{\uparrow}$ levels are primarily occupied for the $\mathcal{B}$ site.
(ii)  The crystal field and exchange splitting of the t$_{2g}$($d_{xy}$,$d_{yz}$ and $d_{xz}$) and e$_{g}$ ($d_{z^2}$ and $d_{x^2-y^2}$)
 levels of Fe substituted (at A site) CoCr$_2$O$_4$ are analysed through  Energy Level Diagram. The level splitting
of only the substituted atoms (Fe for A site and Co$_{O}$ for B site) are shown.The $t_{2g}$ levels are marked purple and the $e_{g}$ levels are marked green.}
\end{figure}
Upon substitution by Fe at A site, the exchange and crystal field splitting of the parent atoms(Co$_{T}$ at A and Cr at B site)
located in the far neighbourhood of the sites of substitutions remain much like the ones of the pristine compound. In the
near neighbourhood of the substituted sites, the densities of states show traces of broken degeneracy (Not shown here). This must be
the effect of structural distortion. Of particular interest in this context is the densities of states for the
displaced Co atom at B site (Co$_{O}$). The d$_{z^{2}}\uparrow$ level for this atom overlaps with the corresponding
d$_{z^{2}}$ level of the Fe atom at A site as shown in Figure 4. As a result the levels show splitting about the Fermi level.
This indicates that the Fe-Co$_{o}$ d$_{z^{2}}$ orbitals hybridise to form molecular orbitals. These split states are
represented by the broken lines in the energy level diagram Figure 3(ii). It is interesting to note that the degeneracy in the
e$_{g}$ states present in the down ($\downarrow$) spin channel of the pristine compound (due to T$_{d}$ symmetry) is broken
due to Fe substitution. This was already indicated by the increase in $\theta_{T_{d}}$. The non-degenerate states, formed,
overlap with the corresponding oxygen p states as expected. The exchange splitting at the A sites for both e$_{g}$ and 
t$_{2g}$ levels increases due to Fe substitution as is seen from 3$^{rd}$ row of Table \ref{tab2}. This ensures that not more than
one electronic state of the substituted atom is occupied in the minority spin channel. However the Co atom
displaced to the octahedral site has a much lower exchange splitting resulting the Co$_{o}$ to be in a low spin state.
\begin{table}[ht]
\caption{\label{tab2} Substituted site ($\mathcal{A}$ and $\mathcal{B}$), Exchange Splitting of $e_{g}$ and $t_{2g}$ levels ($\Delta_{EX}^{e_g}$ and $\Delta_{EX}^{t_{2g}}$ respectively) and Crystal Field Splitting 
of up and down spin channels ($\Delta_{CF}^{\uparrow}$ and $\Delta_{CF}^{\downarrow}$ respectively) for different Species 
(the first two rows are for parent compound where $Co$ atom is at  $\mathcal{A}$ site and $Cr$ atom at  $\mathcal{B}$,
 next two rows are for the substituted Fe at $\mathcal{A}$ site and the Co$_{O}$ atom at $\mathcal{B}$ 
site; the last row is for substituted Fe at $\mathcal{B}$ site).}
\begin{tabular}{lccccc}
\hline
\hline
Site & Species &  $\Delta_{EX}^{e_g}$ & $\Delta_{EX}^{t_{2g}}$ & $\Delta_{CF}^{\uparrow}$  & $\Delta_{CF}^{\downarrow} $\\
   &  & (eV) & (eV) & (eV) & (eV) \\  
\hline
\hline
$\mathcal{A}$ & Co & 1.50 & 2.70 & 1.20 & 0.0\\
$\mathcal{B}$ & Cr & 1.30 & 3.20 & 3.20 & 1.3\\
\hline
$\mathcal{A}$ & Fe & 2.93 & 3.15 & 0.9 & 0.68\\
$\mathcal{B}$ & Co & 0.57 & 0.3 & 1.33 & 1.6\\
\hline
$\mathcal{B}$ & Fe & 2.85 & 3.55 & 2.5 & 1.8\\
\hline
\end{tabular}
\end{table}
Inclusion of the electron-electron correlation through the Coulomb parameter $U$ brings in drastic changes in
the densities of states and subsequently in the exchange splitting. A high spin state of the displaced Co at B site emerges with 
a magnetic moment about 3 $\mu_{B}$ similar to the value at the original tetrahedral site.  Moreover, the direction of the magnetic moment of this
displaced Co is aligned with the Fe moment (Table \ref{tabone:table1}), which was anti-aligned in pure GGA calculations(Fig. \ref{fig:energy-level-A}(ii)). This indicates the failure of 
GGA in describing properly the electronic structure of oxides and the necessity of adding strong Coulomb interactions.
\begin{figure}[h]
\includegraphics[scale=0.4]{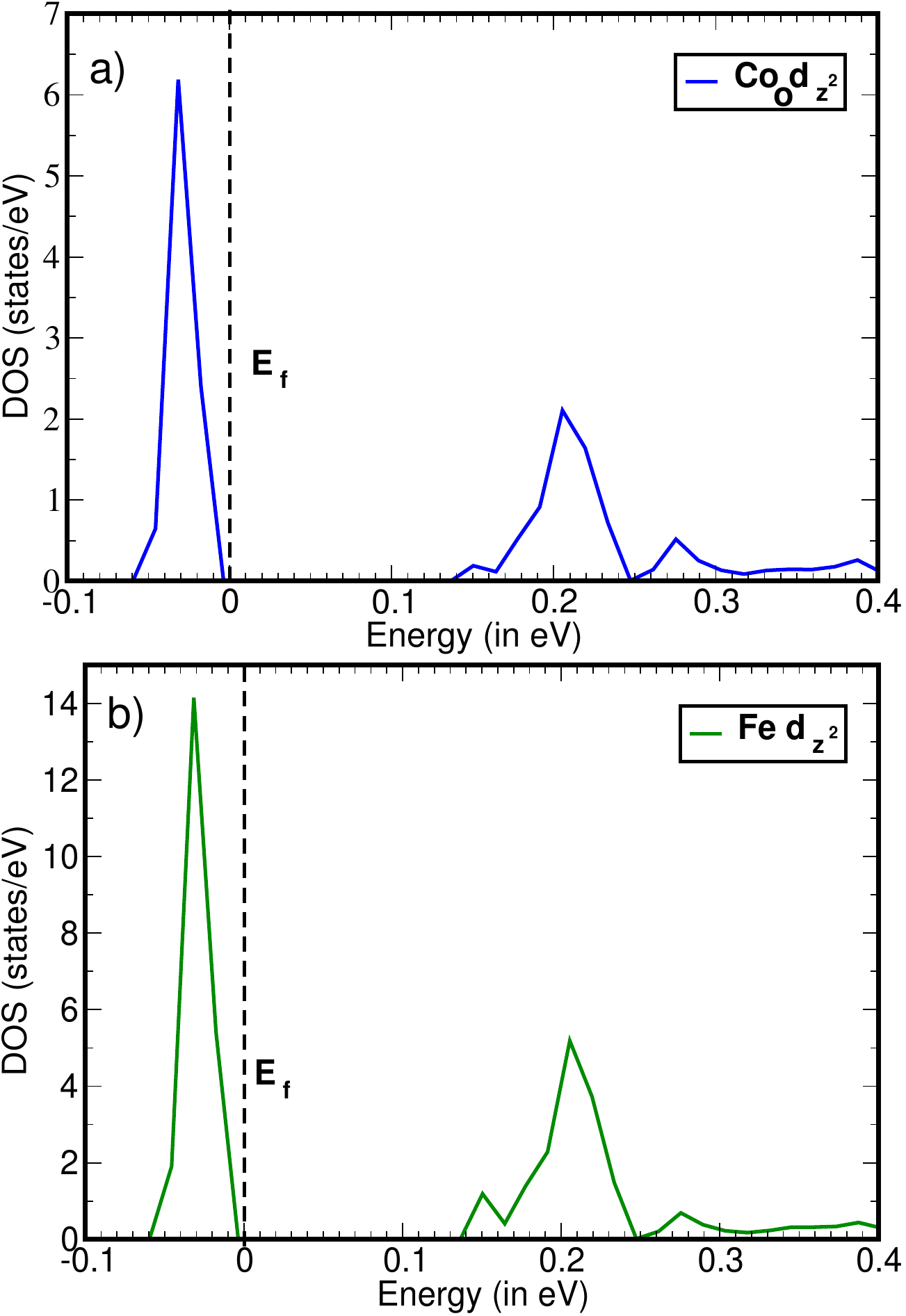}
\caption{\label{fig4} The Co$_{O}$ and Fe densities of states for the d$_{z^{2}}$ orbitals when Fe 
is substituted at A site of CoCr$_{2}$O$_{4}$. The hybridisations show formation of molecular orbitals.
The calculations are for $U_{eff}=0$.}  
\end{figure}
\begin{figure}[h]
\includegraphics[scale=0.35]{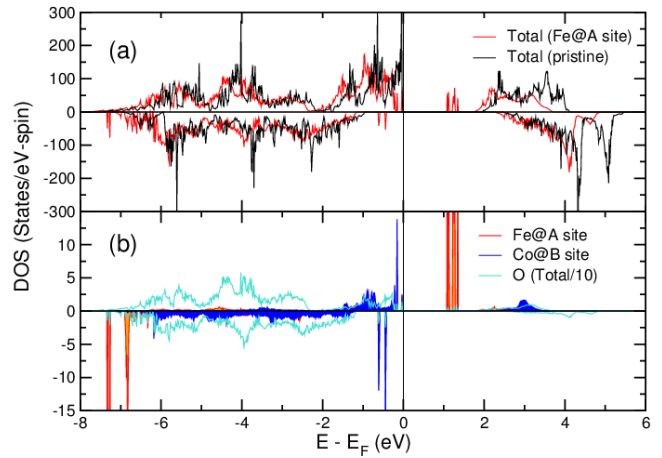}
\caption{\label{fig:DOSA} (top) Total densities of states for pristine CoCr$_{2}$O$_{4}$ and for a single $Fe$ substituted at 
A site in CoCr$_2$O$_4$. (bottom) The atom projected
densities of states for Fe, Co$_{o}$ and oxygen atoms for single Fe substituted
at A site in CoCr$_{2}$O$_{4}$.}` 
\end{figure}
In Figure 5, we compare the 
densities of states of the pristine and the Fe substituted (at A site) CoCr$_{2}$O$_{4}$ along with the
atom projected DOSs. The results for pristine CoCr$_{2}$O$_{4}$ (Figure 5 (top)) qualitatively agree
with previously published one \cite{ederer}. Upon Fe substitution, one can see significant changes in
the densities of states (Figure 5 (top)). New peaks at 1 eV above Fermi level in the up ($\uparrow$) band, at
0.45 eV, 0.65 eV, 6.8 eV and 7.4 eV below Fermi level in the down ($\downarrow$) band appear. The inclusion
of electron-electron correlation pushes the Co$_{o}$ and Fe d$_{z^{2}}$ states, which had formed molecular orbitals
for U=0 case to higher energies. These are the new states appearing in the up band at around 1 eV above Fermi
level. This leads to the correct charge states of +2 and +3 for Co$_{o}$ and Fe atoms respectively.
The peaks at 6.8 eV and 7.4 eV below Fermi level, in the down band, are the contributions mostly from the e$_{g}$ and
t$_{2g}$ states of Fe respectively. The peak at -7.4 eV in the down band results from the
hybridisation of Fe t$_{2g}$ with Co$_{O}$ e$_{g}$ states via adjoining oxygen 2p states and correspond to
Fe-O-Co$_{O}$ superexchange. The peak at -6.8 eV in the down band, similarly, results from hybridisation of Fe
t$_{2g}$ and Co$_{O}$ t$_{2g}$ states via oxygen. The prominent peak at -6.15 eV in the down band results from
hybridisations between Co$_{O}$ e$_{g}$ and Co$_{T}$ t$_{2g}$ via oxygen 2p states, that is, originating via
Co$_{T}$-O-Co$_{O}$ superexchange. The peaks at -0.45 eV and -0.65 eV in the down band arises solely due to
Co$_{O}$ e$_{g}$ states arising due to electron-electron correlation. Such re-distribution of states increases the exchange splitting
at Co$_{O}$ site and helps to achieve the high spin state.
\subsubsection{Fe substitution at B site}
\begin{figure}[ht]
\includegraphics[scale=0.5]{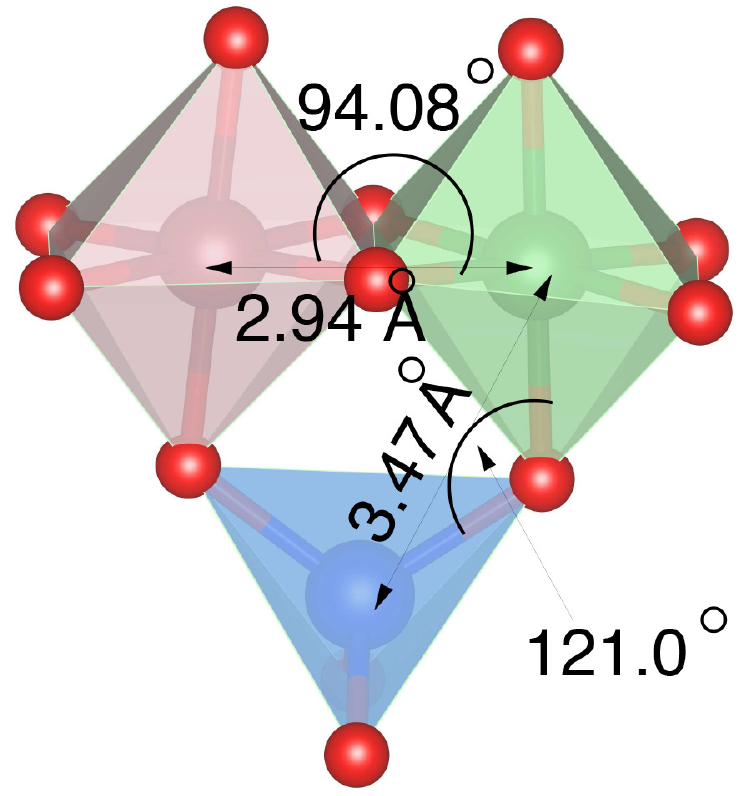}
\caption{\label{fig:b-site} A part of near neighbourhood of the octahedrally coordinated $Fe$ atom 
is shown along with various bond angles and bond lengths for single Fe
substituted at B site in CoCr$_{2}$O$_{4}$. The Co atoms are shaded blue, the Fe atoms
green, the Cr atoms pink and the
O atoms red.}
\end{figure}
The Fe moments prefer to align anti-parallel to those of Cr when Fe is substituted at B site. The total energy
of this configuration is lower by an amount of 413 meV per Fe atom than the minimum energy configuration in case of
Fe substitution at A site. Thus, for this concentration of Fe substitution, the B site substitution is preferred and the
system is a normal spinel. This is in agreement with the experimental observation\cite{pss13} that the inversion starts occurring
for $x>0.05$. In this case too, all the magnetic components are in high spin states and the individual moments do not
change from those in the case of substitution at A site.
\begin{figure}[h]
\includegraphics[scale=0.33]{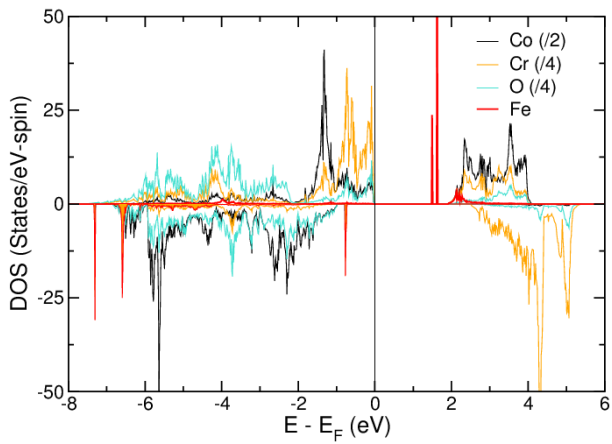}
\caption{\label{fig:DOSB}  Atom projected densities of states for a single $Fe$ substituted at B sublattice in CoCr$_2$O$_4$.}
\end{figure}
In Figure 6, we show the structural distortions around the Fe site and the relevant structural parameters along with the
overall deviations from the octahedral symmetry due to Fe substitution at the octahedral site. It is observed that 
the octahedron around Fe expands, the reason being the substitution of Cr by an atom of larger size. The deviation from
the octahedral symmetry is almost same as that in case of the pristine compound as is seen by comparing $\theta_{O_{h}}$ ($\sim 5.34^{0}$ for the pristine compound, while $\sim 5.27^{0}$ for Fe impurity site in the doped compound) in
both cases. The Fe-O-Co and Fe-O-Cr angles remain almost at the values of the Co-O-Cr and Cr-O-Cr angles in pristine
compound. Even the bond lengths change only slightly from those in the CoCr$_{2}$O$_{4}$. Thus we do not expect to see changes in the electronic 
structure as substantial(in comparison with CoCr$_{2}$O$_{4}$) 
as was observed in case of
substitution at A site.

\begin{figure}[h]
\includegraphics[width=9cm,  keepaspectratio]{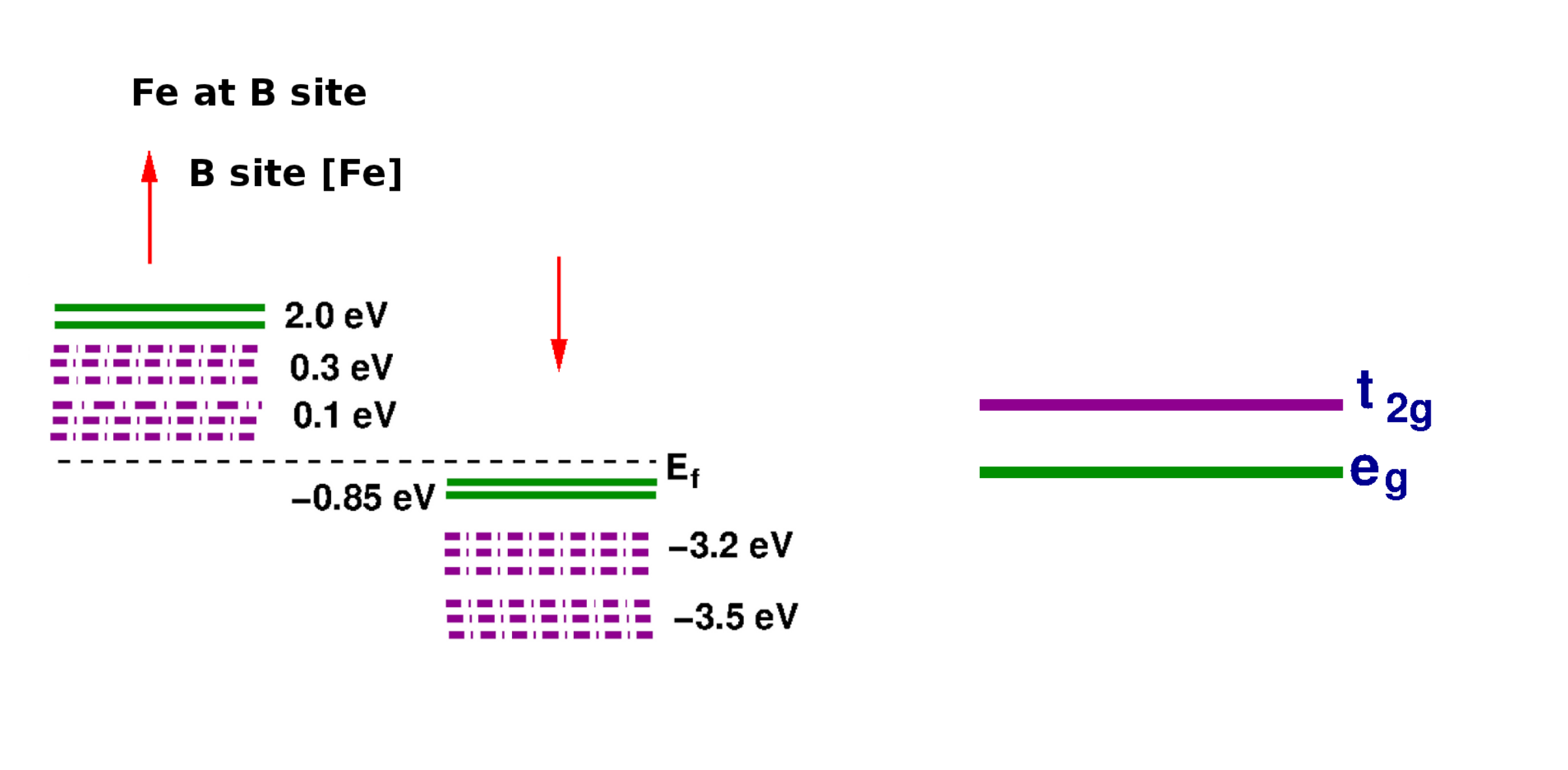}
\caption{\label{fig:energy-level-B}The energy level diagram showing the crystal field and exchange splitting of the t$_{2g}$ and e$_{g}$ 
levels of Fe substituted at B site. The $t_{2g}$ levels are marked purple and the $e_{g}$
levels are marked green.}
\end{figure}

The last row of Table \ref{tab2} lists the crystal field splitting and exchange splitting for the substitution at B site
(without including the effect of electron-electron correlation). Unlike Co substitution at B site, substitution by Fe
enhances the exchange splitting of e$_{g}$ and t$_{2g}$ levels. This ensures that not more than one electronic state of
the substitute is occupied in the minority spin channel. Thus, it appears that the Fe spin state is unaffected by the
symmetry of the site where it is substituted into. In order to understand the impact on the electronic structure, we
once again take recourse to the energy level diagram for the Fe atom in Figure 8. Although the deviations from the
$O_{h}$ symmetry is same as that of the CoCr$_{2}$O$_{4}$, both the Fe t$_{2g}^{\uparrow}$ and t$_{2g}^{\downarrow}$ levels
show splitting. These split states are shown by broken lines in Figure 8. These split levels overlap with the 
adjoining oxygen. Interestingly, the 3-fold and 2-fold degeneracy of the t$_{2g}$ and e$_{g}$ states, respectively, is
maintained at B site even after Fe substitution. In the distant neighbourhood of Fe sites, the situations for Co atoms
at A site and Cr atoms at B sites remain quite similar to that of the pristine CoCr$_{2}$O$_{4}$. In the nearest 
neighbourhood of Fe site, there are 6 octahedrally and 6 tetrahedrally coordinated sites. Both these sites show
reduced symmetry due to Fe substitution and as a result there is a slight loss of degeneracy in the t$_{2g}$ states.

Figure \ref{fig:DOSB} shows the DOSs for a single Fe atom substituted at a B site. It is observed that 
the inclusion of the electron-electron correlation affects the densities of states of Fe, with prominent effects
on the t$_{2g}$ states. The split t$_{2g}^{\uparrow}$ states near the Fermi level are now pushed deeper into the
conduction band while the e$_{g}^{\uparrow}$ states are unaffected. Similarly the e$_{g}^{\downarrow}$
states at -0.85 eV remain unaffected. The electron electron correlation pushes the split t$_{2g}^{\downarrow}$ bands
deeper into the valence band at -7.25 eV. However, the peaks in the total densities of states  are not due to
Fe alone. The peak at -7.25 eV in the down band arises out of hybridisation of Fe t$_{2g}$ and Co e$_{g}$ orbitals through
oxygen 2p. The two peaks at -0.85 eV and -6.5 eV arises out of hybridisations of Fe e$_{g}$ with Co t$_{2g}$ through oxygen 2p.
The peak at -3.9 eV arises out of hybridisation of Fe and Cr e$_{g}$ orbitals, once again via oxygen p orbitals.

\begin{figure}[ht]
\includegraphics[width=8cm,  keepaspectratio]{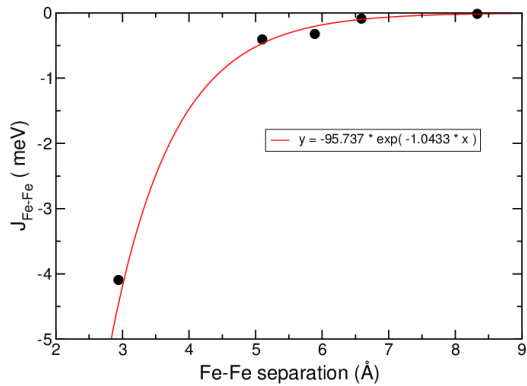}
\caption{\label{fig:dis-vs-en} Interatomic exchange interactions between Fe atoms substituted at B sublattice as a function of Fe-Fe distance. The data points have been fitted (shown as a red line) with
an exponential function as shown in the figure.}
\end{figure}

Thus, we see that the effects of Fe substitution at B site is less dramatic than that at A site. The impact of local structural
distortion is more severe on the electronic structure of Fe substituted CoCr$_{2}$O$_{4}$ at A site. For the case of Fe
substitution at A site, the two Co atoms having different symmetry associated with their sites, bring in additional dimensions
to the impact of substitutions. It is, therefore, expected that the exchange interactions associated with the A site and B sites
would be quite different depending upon the site of Fe substitution. This,in turn, would affect the LKDM parameter $u$ and 
subsequently the occurrence of non collinear states. 
\subsection{Co(Cr$_{0.9375}$Fe$_{0.0625}$)$_{2}$O$_{4}$}
In a 112 atom cell, this composition is modelled by replacing two Cr atoms by two Fe atoms. The distance between two Fe atoms can be 
2.78, 5.1, 5.89, 6.83 and 8.33 \AA. We found that the relative spin orientations of the constituent magnetic atoms
change depending upon the Fe-Fe distance.
In Fig \ref{fig:dis-vs-en}, the interatomic exchange interactions between Fe atoms are shown as a function of Fe-Fe distance. These have been obtained from total energies for minimum energy spin configurations. As seen, all the exchange interactions are antiferromagnetic in nature. This 90$^{\circ}$ antiferromagnetic superexchange interaction follows from Goodenough-Kanamori rules. \cite{goodenough}
Also, we have found that the Fe atoms prefer to stay far away from each other and that the substituted Fe
prefer to align the spins anti-parallel to the Cr spins. As expected, the exchange interactions decay exponentially with distance. This basically indicates that for a low concentration and a homogeneous distribution of Fe atoms, one should not expect a prominent contribution of exchange from Fe atoms in transition temperatures.

The above discussion is for Fe substitution at B sites. We also checked the possibilities of Fe substitution at A sites. Since 
substitution of both Fe atoms at A sites would mean complete inversion, we only looked into the case when one of the two Fe
atoms are substituted into A sites. We found that this configuration lies at an energy 185 meV per Fe atom higher than the
configuration when both Fe occupy B sites. We have earlier reported that for $x=0.03125$, this difference is more than
400 meV per Fe atom. This result, thus, shows that as Fe concentration increases, the tendency for the Fe atoms to occupy
A sites increases, in agreement with experiments \cite{pss13}.

\section{Conclusions}
Using density functional theory based calculations, we have carried out systematic investigations into the 
structural and magnetic properties of Co(Cr$_{1-x}$Fe$_{x}$)$_{2}$O$_{4}$ for low values of $x$ by successive replacement of Cr atoms. 
We have investigated the effects of sublattice occupancies and concentrations on the above mentioned properties. Our results 
show that the substitution of Fe at the tetrahedral sites gives rise
to Co at both tetrahedral and octahedral sites, which behave very differently and produce significant effects on the structural
parameters. The substantial deviation from the tetrahedral symmetry results in loss of degeneracy of the $d$ orbitals of Fe and
Co atoms occupying octahedral sites resulting in a lowering of the magnetic exchange splitting of Co atoms at octahedral positions.
The inclusion of the electron electron correlation enhances the magnetic exchange splitting, thus affecting the spin states of the
transition metal cations. Substitution of Fe at the octahedral site, on the other hand, does not produce such interesting effects, and
the loss of symmetry of the $d$ orbitals as a result of local structural distortion is minimal. Although at very low values of $x$,
the substitution of Fe at B sites are energetically preferable, a tendency towards "inversion" is clearly observed with increasing
$x$, in confirmation with experiments. The magnetic exchange interaction between Fe atoms substituted at B sites decays exponentially with Fe-Fe distance. The antiferromagnetic exchange interactions among Fe atoms along with other magnetic species may give rise to the possibility of frustrated exchange and hence
non-collinear magnetic structures. This should be investigated in detail in future.

\section{Acknowledgments}
SG and BS acknowledge the grant from Swedish Research council VR-SIDA. DD and SG acknowledge the computation facilities from C-DAC, Pune, India
and from Department of Physics, IIT Guwahati funded under the FIST programme of DST, India
.

\end{document}